\documentclass[acmsmall]{acmart} % LBW submission is anonymous
\AtBeginDocument{%
  \providecommand\BibTeX{{%
    \normalfont B\kern-0.5em{\scshape i\kern-0.25em b}\kern-0.8em\TeX}}}

\copyrightyear{2023}
\acmYear{2023}
\setcopyright{rightsretained}
\acmConference[CHI EA '23]{Extended Abstracts of the 2023 CHI Conference on Human Factors in Computing Systems}{April 23--28, 2023}{Hamburg, Germany}
\acmBooktitle{Extended Abstracts of the 2023 CHI Conference on Human Factors in Computing Systems (CHI EA '23), April 23--28, 2023, Hamburg, Germany}\acmDOI{10.1145/3544549.3585841}
\acmISBN{978-1-4503-9422-2/23/04}

\usepackage{enumitem}
\begin{document}

\title[ARctic Escape: A Co-Located Augmented Reality Escape Room]{ARctic Escape: Promoting Social Connection, Teamwork, and Collaboration Using a Co-Located Augmented Reality Escape Room}

\author{Theodore Knoll}
\email{taknoll@princeton.edu}
% \author{Andrés Monroy-Hernández}
% \email{andresmh@princeton.edu }
% \author{Amna Liaqat}
% \email{al0910@princeton.edu}
\affiliation{%
  \institution{Princeton University}
  \city{Princeton}
  \state{New Jersey}
  \country{USA}
  \postcode{08540}
}

\author{Amna Liaqat}
\email{al0910@princeton.edu}
\affiliation{%
  \institution{Princeton University}
  \city{Princeton}
  \state{New Jersey}
  \country{USA}
  \postcode{08540}
}

\author{Andrés Monroy-Hernández}
\email{andresmh@princeton.edu }
\affiliation{%
  \institution{Princeton University}
  \city{Princeton}
  \state{New Jersey}
  \country{USA}
  \postcode{08540}
}
\renewcommand{\shortauthors}{Knoll et al.}

\begin{abstract}
  We present ARctic Escape, a co-located augmented reality (AR) escape room designed to promote collaboration between dyads through play. While physical escape rooms provide groups with fun, social experiences, they require a gameplay venue, props, and a game master, all of which detract from their ease of access. Existing AR escape rooms demonstrate that AR can make escape room experiences easier to access. Still, many AR escape rooms are single-player and therefore fail to maintain the social and collaborative elements of their physical counterparts. This paper presents ARctic Escape, a two-person AR escape room with clues emphasizing player interaction and teamwork. We evaluated ARctic Escape by conducting semi-structured interviews with four dyads to learn about participants’ interpersonal dynamics and experiences during gameplay. We found that participants thought the experience was fun, collaborative, promoted discussion, and inspired new social dynamics, but sometimes the escape room's reliance on virtual content was disorienting.
\end{abstract}



\begin{teaserfigure}
  \includegraphics[width=\textwidth]{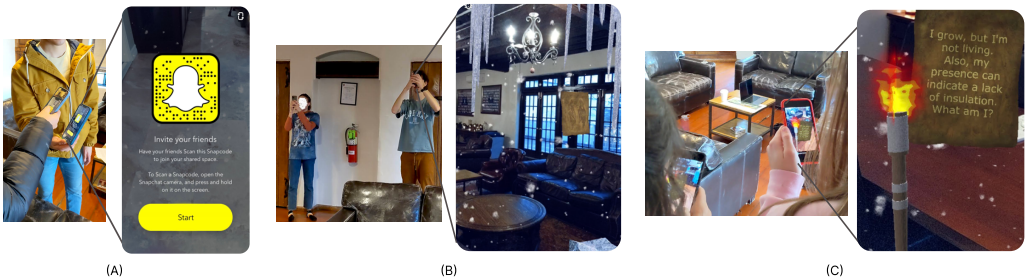}
  \caption{Images during play of the ARctic Escape escape room. (A) After scanning their surrounding space, a player uses a QR-like code to share the escape room with their partner and let them join the game. (B) Augmented reality transforms the space into an escape room with icicles, snow, and other props like an old parchment. (C) Players collaborate by discussing their ideas to solve puzzles and escape the room.}
  \Description{Images during play of the ARctic Escape escape room. (A) After scanning their surrounding space, a player uses a QR-like code to share the escape room with their partner and let them join the game. (B) Augmented reality transforms the space into an escape room with icicles, snow, and other props like an old parchment. (C) Players collaborate by discussing their ideas to solve puzzles and escape the room.}
  \label{fig:teaser}
\end{teaserfigure}


\maketitle

\section{Introduction}
An escape room is an interactive game where players collaborate to discover information about their environment to accomplish a shared goal. Usually, the goal is to "escape" from a room that players imagine is locked shut by finding objects hidden within their surroundings and by solving puzzles \cite{nicholson_peeking_2015}. Although some single-player escape room experiences exist \cite{adver2play_scriptum_2019, afternoon_apps_inc_arias_2022, next-gen_escape_can_2021}, escape rooms tend to be team-oriented experiences that reward players for collaborating and communicating their ideas and discoveries to one another \cite{nicholson_peeking_2015, pan_collaboration_2017}. As a testament to the strength of escape rooms’ ability to foster a social environment, educators and corporations alike have utilized escape room experiences as team-building exercises to train and debrief players on their collaboration and communication skills within a variety of disciplines, ranging from nursing, dermatology, and medicine to investment banking \cite{zhang_trapped_2018, valdes_impact_2021, kelly_adventures_2018, guckian_exploring_2019}.

Escape rooms today typically require a physical venue and physical props. Additionally, they rely on game masters who provide the rules, context, and hints when players get stuck \cite{nicholson_peeking_2015}. Furthermore, like movie theaters or other entertainment sites, physical escape rooms charge an entry fee \cite{nicholson_peeking_2015} and require people to travel to specific venues during specific times. These physical constraints and costs make it harder for everyone to enjoy escape rooms. 

Augmented reality (AR) escape rooms provide an opportunity to facilitate access to escape room experiences while maintaining the valuable affordances of physical co-location by using virtual elements and programmed logic to replace the need for a physical venue, props, and a game master \cite{adver2play_scriptum_2019, afternoon_apps_inc_arias_2022, hsu-chan_lets_2022, tzima_revealing_2021}. Mobile AR uses a camera to display the physical world on a screen and “augments” the displayed world by incorporating virtual artifacts to the input received by the camera \cite{kipper_augmented_2012}. AR escape rooms use virtual artifacts to transform a player’s surrounding environment into an escape room, removing the need for a physical venue and props. Because technology drives the AR escape room experience, developers can incorporate features into the AR escape room experience, like a textual storyline and a hint button, to reduce the need for a game master \cite{adver2play_scriptum_2019}. 

Despite AR escape rooms’ ability to mitigate the ease of access challenges of physical escape rooms, few AR escape rooms exist that both allow anyone with a smartphone to experience an escape room and maintain the social and collaborative elements of physical escape rooms. To address this, we created ARctic Escape, a two-person, co-located AR escape room designed to promote social connection, teamwork, and collaboration.

We evaluated how ARctic Escape’s design and implementation contributed to (1) participants’ interpersonal dynamics with their partner during gameplay and (2) the participants’ experiences within the escape room. We found that participants thought the experience was fun, collaborative, promoted discussion, and inspired new social dynamics, but that sometimes the escape room was disorienting because it was composed solely of virtual content and participants could not view the environment without a device.

\section{Related Work}
There are AR escape rooms that create fun, easy-to-access experiences for people with smartphones capable of running AR applications. Two examples of particularly easy-to-access mobile AR escape room games are \emph{Scriptum AR Escape Room} and \emph{ARia's Legacy}. Both apps are free on the Google Play Store and require limited physical space to play, with \emph{Scriptum} advertising playability in less than three square meters of free space \cite{adver2play_scriptum_2019, afternoon_apps_inc_arias_2022}. Once a user begins playing \emph{Scriptum} or \emph{ARia's Legacy}, their surroundings are transformed on-screen into an escape room, complete with virtual props and buttons to provide players with preprogrammed hints should they need assistance while completing the escape room. In this way, these escape rooms are very convenient to play, because they don’t require a physical venue, props, or a game master. However, both \emph{Scriptum AR Escape Room} and \emph{ARia's Legacy} are single-player experiences, and therefore do not possess the collaborative environment of physical escape rooms we seek to foster.

There are also multiplayer AR escape rooms that succeed in creating collaborative environments, such as \emph{MillSecret}, \emph{AR-Escape}, and \emph{AScapeD}. \emph{MillSecret} lets players collaboratively explore a historical landscape to complete a mix of virtual and physical puzzles, and learn more about the cultural heritage of a former watermill site \cite{tzima_revealing_2021}. Although \emph{MillSecret} succeeds in creating a collaborative AR escape room experience, it can only be played in a specific physical location. \emph{MillSecret} requires a game master, so the escape room does not capitalize on the convenience benefits offered by AR to the escape room experience. Likewise, \emph{AR-Escape} relies on physical props to trigger AR visualizations and progress gameplay, and while players enjoyed interacting with these props, they also made \emph{AR-Escape} harder to access \cite{plecher_david_designing_2020}. Meanwhile, \emph{AScapeD} is a very easy to access experience and only requires AR-capable devices to play, and demonstrates AR escape rooms' high efficacy in promoting socialization between children with ASD and their peers. While prototyping \emph{AScapeD}, Terlouw et al. identify that AR-driven asymmetric information can lead to social interactions between players, but steer away from exploring this further due to observed power imbalances \cite{terlouw_development_2021}. 

We wanted to learn more about how AR can shape collaboration using techniques like asymmetric information, and we designed ARctic Escape to make it easier for two people to play anywhere with only their smartphones. We had ease of access and collaboration in mind, pioneering a co-located AR escape room to promote social connection, teamwork, and collaboration, while trying to make the experience easy to access by only requiring an AR-capable mobile device and a teammate for gameplay. This work provides insights that supplement a body of exciting mobile AR research that explores the design space of playful co-located interaction and uses AR to contribute to meaningful shared experiences \cite{dagan_project_2022}.

\section{ARctic Escape System}
\begin{figure}[h]
  \centering
  \includegraphics[width=\linewidth]{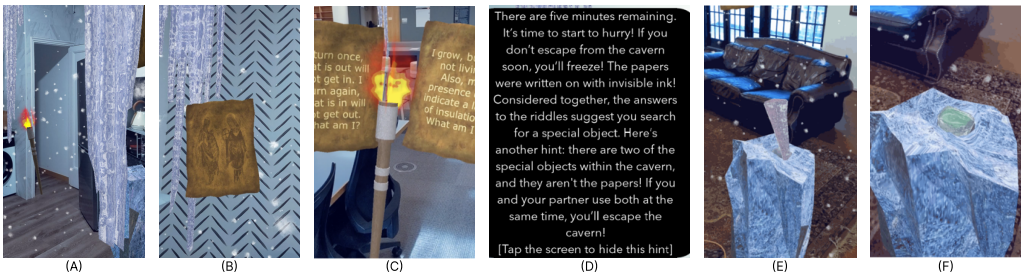}
  \caption{A typical flow of the ARctic Escape as players work to complete the escape room. Note that the clues do not have to be solved in linear order, but players often solve easier clues first. (A) Players begin by exploring the room. (B) Players find parchment, which may have a hidden purpose. (C) Players discover how to read clues on the parchments using a torch. (D) A timed hint appears to help players solve the previous riddles. (E) Players recognize the icicle keys must go into ice keyholes, but fail to properly insert the keys. (F) Players simultaneously insert the icicle keys into keyholes, solving the escape room.}
  \label{fig:Flow of Play}
\end{figure}

ARctic Escape enables co-located players to use their mobile phones to digitally transform their surroundings into an Arctic-themed escape room. The Arctic theme was arbitrarily chosen to provide an immersive escape room experience.

ARctic Escape is playable through Snapchat as a Snapchat Lens\footnote{Snapchat Lenses are AR experiences accessible through Snapchat}. One player can load up the ARctic Escape lens and scan their surroundings using their phone’s camera in a process that takes about 10 seconds. After the room is sufficiently scanned, the player can use a link or QR code to share the escape room with their partner (Fig. \ref{fig:teaser}A). When the game begins, the virtual escape room environment appears around them and players can start exploring in search of clues (Fig. \ref{fig:teaser}B).

We implemented ARctic Escape using Lens Studio\footnote{Lens Studio is a Snap AR development tool for creating augmented reality experiences accessible on Snapchat} and Connected Lens technology\footnote{Connected Lens technology is a Snap AR interface that enables multiple participants to share an AR experience}, in conjunction with JavaScript to build out the game mechanics. Images for the project came from Unsplash\footnote{Unsplash is a free photo website}, and we acquired most of the 3D assets from \textit{TurboSquid}\footnote{\textit{TurboSquid} is an online 3D model vendor with free and paid digital assets}. We also created custom assets for the game using Blender\footnote{Blender is a free, open-source 3D creation suite}.

\begin{figure}[h]
  \centering
  \includegraphics[width=\linewidth]{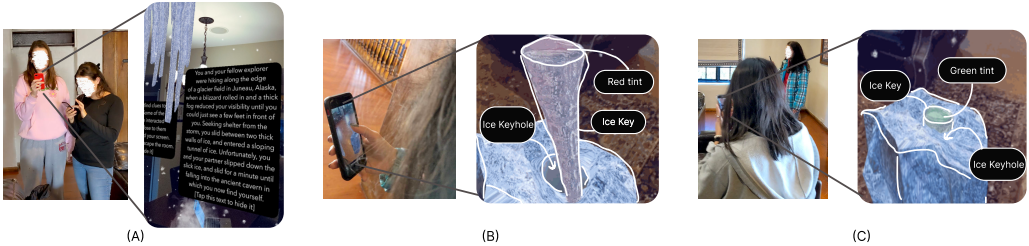}
  \caption{Features of the escape room that let the game be played anywhere (A) and promote collaboration (B and C). (A) Players enter the escape room and encounter lore for the game and a description of the controls. (B) A player independently inserts an icicle key into one of two ice keyholes within the room. Because their teammate is not simultaneously inserting another key into the other keyhole, the icicle turns red and vacates it. (C) Both players simultaneously insert icicle keys into unique ice keyholes (only one keyhole is depicted). Because the players collaborate through synchronized insertion, the icicle turns green and fills the keyhole, triggering the end sequence.}
  \label{fig:playable-anywhere}
\end{figure}

We designed ARctic Escape as a convenient, easy-to-access escape room experience that promotes collaboration using the following techniques:

\begin{description}
\item[Convenient and easy to access play.] The game lasts at most 20 minutes. Players can enjoy ARctic Escape anywhere they can scan a space of approximately three square meters. This space does not have to be void of other objects.

\item[Interface reduces the need for a game master.] When players load the game, they encounter a text block that describes the lore of how players got “trapped” within ARctic escape. Another text block explains that players tap on virtual objects displayed to interact with them (Fig. \ref{fig:playable-anywhere}A). We chose to have on-screen taps control object interactions because hand-tracking behaved inconsistently during our development. At five-minute intervals, the game displays hints about the escape room to both players. These hints progressively reveal more information about the escape room and guide players toward completion.

\item[Collaboration-promoting experience.] ARctic Escape is a co-located game, and the shared space allows players to communicate their thoughts and discoveries to one another as they explore the room and work to escape. Objects in the room interact with one another, making for exciting shareable experiences. For example, two pieces of parchment behave as though they were written on with invisible ink, and reveal riddles on them when they are brought close to a torch (Fig. \ref{fig:teaser}C). The system does not provide the answers to the riddles, so players talk through their ideas to determine the riddles’ solutions.

\item[More intimate bonding experience.] The escape room is designed for two players to ensure a more personal gameplay experience. The two-person experience also offers the benefit of making it easier to observe and analyze the social interactions between players, in comparison to larger groups that may have more complex social dynamics. Additionally, having a set number of players helped us focus the gameplay and create clues that encourage collaboration between all players. For example, we created a key/keyhole puzzle: there are two icicle keys within the room, and two ice keyholes spread apart from one another. To complete ARctic Escape, players must each simultaneously insert an icicle key into a unique ice keyhole (Fig. \ref{fig:playable-anywhere}C). This means that successful play requires effective communication.

\item[Build on information asymmetry.] The game displays different virtual content on each player's screen if players do not collaborate while solving the key/keyhole puzzle described above. Specifically, suppose Player A inserts an icicle key into a keyhole without Player B inserting a key into the other keyhole. In that case, all icicles will temporarily turn red on Player A’s screen, but the icicles will not change color on Player B’s screen (Fig. \ref{fig:playable-anywhere}B). Player A must inform Player B of the color change for the players to recognize that solving the puzzle might require working together.\label{asymmetric-visual-information}
\end{description}
\section{Evaluation}
We recruited four dyads\footnote{The term dyad refers to a group of two people} (eight total participants) of university students to play ARctic Escape indoors. All participants were American citizens. The average age was 20.875 (SD = 0.78). Five participants identified as women, and three as men. Participants identified as White (4), Asian-American (1), Asian (1), and White/Asian (2). When a dyad arrived at the location of the study, we instructed them how to open up the ARctic Escape Lens on Snapchat and walked them through scanning the room. Then they began the game, and played until completion of the escape room or 20 minutes had passed, whichever came first. Following the game, we conducted a semi-structured interview with the dyad to learn about their experiences in the escape room and to understand how it affected teams' interpersonal dynamics during gameplay (see appendix).

\section{Results}
ARctic Escape successfully created an environment where players could socialize, collaborate, and have fun. Below, we group results from the study into three categories, focusing on the shared social experience, factors that influenced communication, and AR’s immersion and participants’ experiences within physical and virtual space.\footnote{Quotes used have been lightly edited for readability by removing filler words}

\subsection{Collaboration and teamwork came easily to participants, and sharing the experience with a friend increased enjoyment}
Upon starting a game, all of the participants immediately began to explore the AR environment and share their discoveries with one another. When describing how their team worked together, P3 said, \textit{“we were talking throughout it, which was helpful,”} and that they would \textit{“bounce ideas off one another”}. P1 said that during their game alongside P2, \textit{“there was some intuitive teamwork without communication.”} Additionally, P7 said, \textit{“I just talked to [P8] like I always talk to [P8]. I made little jokes and stuff.”}

Participants also described how sharing the experience with their partner made it more fun. P6 said,\textit{ “It was a game you can do with your friends. I really liked that. I feel like it wouldn't be as fun if you were just doing it by yourself.”} P2 expressed a similar sentiment, saying that \textit{“I had a lot more fun”} because they shared the experience with P1. All participants said they enjoyed playing ARctic Escape, and even P4, who rated the AR escape room a 3.5/10, the lowest of all participants (M = 7.06, SD = 1.74), said, \textit{“I enjoyed it because it’s kind of cool technology, I’m doing it with a friend.”} P4 described their dissatisfaction with the experience by saying, \textit{"there's an inherent jankiness, or strangeness of using the AR technology,"} and expressed a desire to find tangible objects in the escape room (see \ref{immersion}), but P4 also said, \textit{“escape rooms are supposed to be collaborative. So I think it broke a little bit of the weirdness of the AR thing.”}

\subsection{Co-located play helped participants communicate, and asymmetric visual information led to heightened curiosity and created teaching opportunities}
Playing the game in the same room as one another helped participants communicate freely with their partner. For example, one of the participants mentioned that they \textit{“said everything that popped into [their] head”} (P8). Throughout the game, participants would often look away from their phones and speak while facing their partner. The ability to communicate directly also helped participants to discuss and understand asymmetric visual information (see \S\ref{asymmetric-visual-information}). Upon observing an icicle change color from white to red, the observer almost always informed their teammate. This led the teammate to become curious about what the observer had witnessed, and they asked questions like, \textit{"What?" }(P2), and \textit{"How did you put [the icicle] in?"} (P3). Following this questioning, the observer taught their partner how to achieve the same result. For example, P1 said, \textit{"I put [the icicle] in the [ice keyhole], it turns pink, and floats back up."} Two of the four dyads completed ARctic Escape successfully, and once the color change observer taught their partner how to insert the icicle key into the keyhole, both of the successful groups quickly determined that the asymmetric visual information was a clue that encouraged players to work together.

\subsection{Participants liked the immersion provided by AR and had fun interacting in the space in new ways, but experiences from the physical world did not transfer seamlessly to an AR setting}\label{immersion}
Participants expressed significant interest in the virtual space and appreciated its ability to supplement the physical space with an escape room theme. P4 said that \textit{“having the snow come down added to the atmosphere,”} and P2 stated, \textit{“it was fun to be in a little space with icicles actually hanging in front of you.”} This immersion also changed the way participants moved within the physical world. P5 commented how \textit{“it was fun to be playing the game where I interacted with the normal space in a different way… I would never be walking around crouching, doing weird, strange things in this space. So it was cool to have that recontextualized.”} Participants also compared the use of AR to other escape room mediums. P1 said, \textit{“making it AR versus just having it be a digital click-through game makes it feel more immersive, and more like an actual escape room,”} and P4 said that in comparison to physical escape rooms, AR had a \textit{“cool added advantage, that you could do it in any space.”}

However, participants also had trouble within the escape room because sometimes their experiences in the physical world did not transfer well to the AR escape room. P5 said, \textit{“it was hard to wrap myself around certain things mentally,”} and that \textit{“the space was a bit disorienting,”} because \textit{“I can usually look at the whole space at once and get an idea of where things are in relation to each other… the two ice boxes we were working with felt really disconnected because you couldn’t see them both at the same time.”} Other participants commented that unlike in the physical world, it was difficult to tell when a player was holding an object in-game. Participants also expressed that AR made exploring the room more difficult, and P8 said, \textit{“I don't feel super comfortable with the physics and mechanics of the game.”} Some participants also felt that holding virtual items was unsatisfying due to the lack of tactile stimuli. P4 said, \textit{"even when I'm moving things around, I don't really feel like I'm holding them. Having the physical sensation of finding something, it just feels more real."}

\section{Discussion}
The results of our study draw attention to how the design of a group AR experience can shape how individuals communicate. The results reveal that co-located AR experiences allow individuals to easily exchange information. Therefore, co-location creates an environment well-suited for collaboration and developing social connections. Within ARctic Escape, co-location helped players talk through ideas and work together to solve the puzzles they encountered in the escape room. Beyond the ease that co-location provided to verbal communication, having the ability to look away from the AR experience and directly view their partner's expressions and body language likely helped participants communicate more effectively. By demonstrating co-location's efficacy in supporting communication, these findings extend existing research arguing that mobile AR can effectively support co-located play \cite{dagan_project_2022}. Furthermore, our results suggest that designers can use asymmetric visual information in AR to stimulate discussion and craft new social dynamics. By providing one individual with more information than another, asymmetric visual information creates a power dynamic that may not exist outside of the AR experience, which designers could use to encourage leadership among individuals who typically conform to others' lead.

The study also shows that individuals find joy in shared social activities, and feedback from participants indicated that sharing the AR experience with a partner increased their tolerance for technical issues and awkward game mechanics. These findings suggest that single-player AR escape rooms \cite{adver2play_scriptum_2019, afternoon_apps_inc_arias_2022} may become more enjoyable by increasing the number of players they support. Ideally, game designers can create a smooth and intuitive AR experience that players enjoy, but that is not always possible, and players have different preferences for what they find fun. Fortunately, by providing opportunities within an AR experience for players to interact with one another, our study revealed that players can generate their own fun experiences with their partner, and do not need to rely exclusively on technology to provide them with entertainment.

\section{Limitations}
Our findings come from a small group of participants chosen via convenience sampling, and all participants were US college students, a largely homogeneous population. We asked participants to bring a friend for the study, so all dyads were composed of individuals who already knew and felt comfortable with one another. Additional research is necessary to explore how AR escape rooms influence interpersonal dynamics between strangers.

The AR escape room experience was shorter than the typical physical escape room and had relatively few puzzles for participants to solve to complete the escape room. This could have led players to work together more often than in larger and more complex AR escape rooms. We also experienced some playability issues, such as crashes during gameplay. When these issues occurred, we helped participants rejoin the game, and we used text on a laptop to deliver the hints to teams that they would have received on their phones at five-minute intervals in the absence of the crashes. In this way, we partially served as the game master we sought to replace with technology. However, without crashes, we believe having preprogrammed hints is still an effective way to make escape rooms easier to access. 

\section{Conclusion}
In this paper, we presented ARctic Escape, a co-located AR escape room designed to promote collaboration. Through ARctic Escape, we created an escape room experience that is playable anywhere by anyone with an AR-capable mobile device, removing physical escape rooms' need for a venue, physical props, and a game master. We deployed ARctic Escape and conducted a study with four dyads. In doing so, we added to a growing collection of playful, co-located AR experiences and further demonstrated that co-located AR can offer meaningful opportunities for social connection and shared experiences. From our study, we began to explore how co-location affects communication within AR experiences, and found that designers can use asymmetric visual information to provide users with the opportunity to lead and teach a group, providing evidence that AR escape rooms could serve as an effective medium for novel team and leadership-building experiences. We believe that using AR to allow individuals to experience new social dynamics can increase their social versatility and improve their ability to work well within teams. Moreover, our study revealed that sharing an AR experience with a friend increased the receptivity of users to the experience, even when technical errors occurred. This showcased the potentially enormous value that collaborative and co-located AR experiences could have to help increase the adoption of AR technology in our society. We hope that our AR escape room and study inspire future work that explores how AR shapes social connection and communication.

% \begin{acks}
% \end{acks}

\bibliographystyle{ACM-Reference-Format}
\bibliography{ARctic-Escape-Bib}


\begin{thebibliography}{15}

%%% the Web address.

\ifx \showCODEN    \undefined \def \showCODEN     #1{\unskip}     \fi
\ifx \showDOI      \undefined \def \showDOI       #1{#1}\fi
\ifx \showISBNx    \undefined \def \showISBNx     #1{\unskip}     \fi
\ifx \showISBNxiii \undefined \def \showISBNxiii  #1{\unskip}     \fi
\ifx \showISSN     \undefined \def \showISSN      #1{\unskip}     \fi
\ifx \showLCCN     \undefined \def \showLCCN      #1{\unskip}     \fi
\ifx \shownote     \undefined \def \shownote      #1{#1}          \fi
\ifx \showarticletitle \undefined \def \showarticletitle #1{#1}   \fi
\ifx \showURL      \undefined \def \showURL       {\relax}        \fi
% invisible to TeX
\providecommand\bibfield[2]{#2}
\providecommand\bibinfo[2]{#2}
\providecommand\natexlab[1]{#1}
\providecommand\showeprint[2][]{arXiv:#2}

\bibitem[{Adver2Play}(2019)]%
        {adver2play_scriptum_2019}
\bibfield{author}{\bibinfo{person}{{Adver2Play}}.}
  \bibinfo{year}{2019}\natexlab{}.
\newblock \bibinfo{title}{Scriptum {AR} {Escape} {Room} - {Apps} on {Google}
  {Play}}.
\newblock
\newblock
\urldef\tempurl%
\url{https://play.google.com/store/apps/details?id=com.Adver2Play.Scriptum&hl=en_CA&gl=US}
\showURL{%
\tempurl}


\bibitem[{Afternoon Apps Inc.}(2022)]%
        {afternoon_apps_inc_arias_2022}
\bibfield{author}{\bibinfo{person}{{Afternoon Apps Inc.}}}
  \bibinfo{year}{2022}\natexlab{}.
\newblock \bibinfo{title}{{ARia}'s {Legacy} - {AR} {Escape} {Room}}.
\newblock
\newblock
\urldef\tempurl%
\url{https://play.google.com/store/apps/details?id=com.AfternoonAppsInc.ARia&hl=en_CA&gl=US}
\showURL{%
\tempurl}


\bibitem[Dagan et~al\mbox{.}(2022)]%
        {dagan_project_2022}
\bibfield{author}{\bibinfo{person}{Ella Dagan}, \bibinfo{person}{Ana~María
  Cárdenas~Gasca}, \bibinfo{person}{Ava Robinson}, \bibinfo{person}{Anwar
  Noriega}, \bibinfo{person}{Yu~Jiang Tham}, \bibinfo{person}{Rajan Vaish},
  {and} \bibinfo{person}{Andrés Monroy-Hernández}.}
  \bibinfo{year}{2022}\natexlab{}.
\newblock \showarticletitle{Project {IRL}: {Playful} {Co}-{Located}
  {Interactions} with {Mobile} {Augmented} {Reality}}.
\newblock \bibinfo{journal}{\emph{Proceedings of the ACM on Human-Computer
  Interaction}} \bibinfo{volume}{6}, \bibinfo{number}{CSCW1}
  (\bibinfo{date}{April} \bibinfo{year}{2022}), \bibinfo{pages}{62:1--62:27}.
\newblock
\urldef\tempurl%
\url{https://doi.org/10.1145/3512909}
\showDOI{\tempurl}


\bibitem[Guckian et~al\mbox{.}(2019)]%
        {guckian_exploring_2019}
\bibfield{author}{\bibinfo{person}{J. Guckian}, \bibinfo{person}{A. Sridhar},
  {and} \bibinfo{person}{S.~J. Meggitt}.} \bibinfo{year}{2019}\natexlab{}.
\newblock \showarticletitle{Exploring the perspectives of dermatology
  undergraduates with an escape room game}.
\newblock \bibinfo{journal}{\emph{Clinical and Experimental Dermatology}}
  \bibinfo{volume}{45}, \bibinfo{number}{2} (\bibinfo{date}{July}
  \bibinfo{year}{2019}), \bibinfo{pages}{153--158}.
\newblock
\showISSN{1365-2230}
\urldef\tempurl%
\url{https://doi.org/10.1111/ced.14039}
\showDOI{\tempurl}
\newblock
\shownote{Publisher: John Wiley \& Sons, Ltd}.


\bibitem[Hsu-Chan et~al\mbox{.}(2022)]%
        {hsu-chan_lets_2022}
\bibfield{author}{\bibinfo{person}{Kuo~1 Hsu-Chan}, \bibinfo{person}{Ai-Jou
  Pan 2}, \bibinfo{person}{Lin~3 Cai-Sin}, \bibinfo{person}{Chang 4 1 Center
  of Teacher~Education Chu-Yang}, {and} \bibinfo{person}{National Cheng
  Kung~University Graduate Institute~of Education}.}
  \bibinfo{year}{2022}\natexlab{}.
\newblock \showarticletitle{Let’s {Escape}! {The} {Impact} of a
  {Digital}-{Physical} {Combined} {Escape} {Room} on {Students}’ {Creative}
  {Thinking}, {Learning} {Motivation}, and {Science} {Academic} {Achievement}}.
\newblock  (\bibinfo{year}{2022}), \bibinfo{pages}{615}.
\newblock
\urldef\tempurl%
\url{https://doi.org/10.3390/educsci12090615}
\showDOI{\tempurl}
\newblock
\shownote{Num Pages: 615 Publisher: MDPI AG}.


\bibitem[Kelly(2018)]%
        {kelly_adventures_2018}
\bibfield{author}{\bibinfo{person}{Lindsay Kelly}.}
  \bibinfo{year}{2018}\natexlab{}.
\newblock \showarticletitle{Adventures in teambuilding}.
\newblock \bibinfo{journal}{\emph{Northern Ontario Business}}
  \bibinfo{volume}{38}, \bibinfo{number}{4} (\bibinfo{date}{Feb.}
  \bibinfo{year}{2018}), \bibinfo{pages}{25,28}.
\newblock
\showISSN{07102755}
\urldef\tempurl%
\url{https://www.proquest.com/docview/2042741989/citation/75F36668C0104AFBPQ/1}
\showURL{%
\tempurl}
\newblock
\shownote{Num Pages: 25,28 Place: Sudbury, Canada Publisher: Northern Ontario
  Business Section: Business Travel}.


\bibitem[Kipper and Rampolla(2012)]%
        {kipper_augmented_2012}
\bibfield{author}{\bibinfo{person}{Gregory Kipper} {and}
  \bibinfo{person}{Joseph Rampolla}.} \bibinfo{year}{2012}\natexlab{}.
\newblock \bibinfo{booktitle}{\emph{Augmented {Reality}: {An} {Emerging}
  {Technologies} {Guide} to {AR}}}.
\newblock \bibinfo{publisher}{Elsevier Science \& Technology Books},
  \bibinfo{address}{Saint Louis, UNITED STATES}.
\newblock
\showISBNx{978-1-59749-734-3}
\urldef\tempurl%
\url{http://ebookcentral.proquest.com/lib/princeton/detail.action?docID=1073012}
\showURL{%
\tempurl}


\bibitem[{Next-Gen Escape}(2021)]%
        {next-gen_escape_can_2021}
\bibfield{author}{\bibinfo{person}{{Next-Gen Escape}}.}
  \bibinfo{year}{2021}\natexlab{}.
\newblock \bibinfo{title}{Can you go to an escape room alone? {Solo}
  {Escapes}!}
\newblock
\newblock
\urldef\tempurl%
\url{https://nextgenescape.com/blogs/escape-room-tips-tricks/can-you-go-to-an-escape-room-alone-solo-escapes}
\showURL{%
\tempurl}


\bibitem[Nicholson(2015)]%
        {nicholson_peeking_2015}
\bibfield{author}{\bibinfo{person}{Scott Nicholson}.}
  \bibinfo{year}{2015}\natexlab{}.
\newblock \bibinfo{title}{Peeking {Behind} the {Locked} {Door}: {A} {Survey} of
  {Escape} {Room} {Facilities}}.
\newblock
\newblock
\urldef\tempurl%
\url{https://scottnicholson.com/pubs/erfacwhite.pdf}
\showURL{%
\tempurl}


\bibitem[Pan et~al\mbox{.}(2017)]%
        {pan_collaboration_2017}
\bibfield{author}{\bibinfo{person}{Rui Pan}, \bibinfo{person}{Henry Lo}, {and}
  \bibinfo{person}{Carman Neustaedter}.} \bibinfo{year}{2017}\natexlab{}.
\newblock \showarticletitle{Collaboration, {Awareness}, and {Communication} in
  {Real}-{Life} {Escape} {Rooms}}. In \bibinfo{booktitle}{\emph{Proceedings of
  the 2017 {Conference} on {Designing} {Interactive} {Systems}}}.
  \bibinfo{publisher}{ACM}, \bibinfo{address}{Edinburgh United Kingdom},
  \bibinfo{pages}{1353--1364}.
\newblock
\showISBNx{978-1-4503-4922-2}
\urldef\tempurl%
\url{https://doi.org/10.1145/3064663.3064767}
\showDOI{\tempurl}


\bibitem[Plecher et~al\mbox{.}(2020)]%
        {plecher_david_designing_2020}
\bibfield{author}{\bibinfo{person}{David Plecher}, \bibinfo{person}{Maximilian
  Ludl}, {and} \bibinfo{person}{Gudrun Klinker}.}
  \bibinfo{year}{2020}\natexlab{}.
\newblock \showarticletitle{Designing an {AR}-{Escape}-{Room} with
  {Competitive} and {Cooperative} {Mode}}.
\newblock  (\bibinfo{year}{2020}).
\newblock
\urldef\tempurl%
\url{https://doi.org/10.18420/VRAR2020_30}
\showDOI{\tempurl}
\newblock
\shownote{Publisher: Gesellschaft f{\"u}r Informatik e.V.}.


\bibitem[Terlouw et~al\mbox{.}(2021)]%
        {terlouw_development_2021}
\bibfield{author}{\bibinfo{person}{Gijs Terlouw}, \bibinfo{person}{Derek
  Kuipers}, \bibinfo{person}{Job van~'t Veer}, \bibinfo{person}{Jelle~T Prins},
  {and} \bibinfo{person}{Jean Pierre E~N Pierie}.}
  \bibinfo{year}{2021}\natexlab{}.
\newblock \showarticletitle{The {Development} of an {Escape} {Room}--{Based}
  {Serious} {Game} to {Trigger} {Social} {Interaction} and {Communication}
  {Between} {High}-{Functioning} {Children} {With} {Autism} and {Their}
  {Peers}: {Iterative} {Design} {Approach}}.
\newblock \bibinfo{journal}{\emph{JMIR Serious Games}} \bibinfo{volume}{9},
  \bibinfo{number}{1} (\bibinfo{date}{March} \bibinfo{year}{2021}),
  \bibinfo{pages}{e19765}.
\newblock
\showISSN{2291-9279}
\urldef\tempurl%
\url{https://doi.org/10.2196/19765}
\showDOI{\tempurl}


\bibitem[Tzima et~al\mbox{.}(2021)]%
        {tzima_revealing_2021}
\bibfield{author}{\bibinfo{person}{Stavroula Tzima}, \bibinfo{person}{Georgios
  Styliaras}, {and} \bibinfo{person}{Athanasios Bassounas}.}
  \bibinfo{year}{2021}\natexlab{}.
\newblock \showarticletitle{Revealing {Hidden} {Local} {Cultural} {Heritage}
  through a {Serious} {Escape} {Game} in {Outdoor} {Settings}}.
\newblock \bibinfo{journal}{\emph{Information}} \bibinfo{volume}{12},
  \bibinfo{number}{1} (\bibinfo{year}{2021}), \bibinfo{pages}{10}.
\newblock
\urldef\tempurl%
\url{https://doi.org/10.3390/info12010010}
\showDOI{\tempurl}
\newblock
\shownote{Num Pages: 10 Place: Basel, Switzerland Publisher: MDPI AG}.


\bibitem[Valdes et~al\mbox{.}(2021)]%
        {valdes_impact_2021}
\bibfield{author}{\bibinfo{person}{Beatriz Valdes}, \bibinfo{person}{Mary
  Mckay}, {and} \bibinfo{person}{Jill~S. Sanko}.}
  \bibinfo{year}{2021}\natexlab{}.
\newblock \showarticletitle{The {Impact} of an {Escape} {Room} {Simulation} to
  {Improve} {Nursing} {Teamwork}, {Leadership} and {Communication} {Skills}:
  {A} {Pilot} {Project}}.
\newblock \bibinfo{journal}{\emph{Simulation \& Gaming}} \bibinfo{volume}{52},
  \bibinfo{number}{1} (\bibinfo{date}{Feb.} \bibinfo{year}{2021}),
  \bibinfo{pages}{54--61}.
\newblock
\showISSN{1046-8781}
\urldef\tempurl%
\url{https://doi.org/10.1177/1046878120972738}
\showDOI{\tempurl}
\newblock
\shownote{Publisher: SAGE Publications Inc}.


\bibitem[Zhang et~al\mbox{.}(2018)]%
        {zhang_trapped_2018}
\bibfield{author}{\bibinfo{person}{Xiao~Chi Zhang}, \bibinfo{person}{Hyunjoo
  Lee}, \bibinfo{person}{Carlos Rodriguez}, \bibinfo{person}{Joshua Rudner},
  \bibinfo{person}{Teresa~M Chan}, {and} \bibinfo{person}{Dimitrios
  Papanagnou}.} \bibinfo{year}{2018}\natexlab{}.
\newblock \showarticletitle{Trapped as a {Group}, {Escape} as a {Team}:
  {Applying} {Gamification} to {Incorporate} {Team}-building {Skills} {Through}
  an ‘{Escape} {Room}’ {Experience}}.
\newblock \bibinfo{journal}{\emph{Cureus}} \bibinfo{volume}{10},
  \bibinfo{number}{3} (\bibinfo{date}{March} \bibinfo{year}{2018}),
  \bibinfo{pages}{e2256}.
\newblock
\showISSN{2168-8184}
\urldef\tempurl%
\url{https://doi.org/10.7759/cureus.2256}
\showDOI{\tempurl}


\end{thebibliography}

\appendix
\section{Interview Questions}
Below are the questions that guided our semi=structured interviews with dyads:
\begin{enumerate}
\item Have you done an escape room before?
\item Have you used an AR application before?
\item What were your thoughts on the AR escape room experience?
\item What was easy or difficult about using this application?
\item What are your thoughts about the clues?
\item Rate the level of difficulty of the clues on a scale of 1-10, with 1 meaning the clues were trivial, and 10 meaning the clues were so hard you don’t think anyone could solve it without already knowing the answer. Explain why you chose that score.
\item How did the use of AR affect the escape room experience?
\item Describe your team dynamic within the escape room and while solving clues.
\item Describe how you and your partner communicated during the game.
\item Did any aspects of the AR escape room promote or deter collaboration with your partner? Explain.
\item Rate your experience doing the AR escape room on a scale of 1-10, with 1 meaning you hated the experience, and 10 meaning you loved the experience. Explain why you chose that score.
\item Rate how comfortable you felt during the AR escape room experience on a scale of 1 to 10, with 1 meaning very uncomfortable to 10 meaning very comfortable. What made you feel this way?
\end{enumerate}

\end{document}